\documentclass[pra,aps,twocolumn,twoside,english]{revtex4}

\usepackage{amssymb}
\usepackage{amsmath}
\usepackage{amscd}
\usepackage{babel}
\usepackage{latexsym} 
\usepackage{epsfig}
\usepackage{bbm}

\newtheorem{lemma}{Lemma}
\newtheorem{thm}{Theorem}
\newtheorem{corollary}{Corollary}

\newcommand{\qed}{{\hfill$\Box$}}
\newenvironment{proof}{\noindent \textbf{{Proof~} }}{\qed}

\def\bi{\begin{itemize}}
\def\ei{\end{itemize}}
\def\be{\begin{equation}}
\def\ee{\end{equation}}
\def\bea{\begin{eqnarray}}
\def\eea{\end{eqnarray}}
\def\ben{\begin{eqnarray*}}
\def\een{\end{eqnarray*}}

\def\>{\rangle}
\def\<{\langle}

\def\ra{\rightarrow}

\def\eps{\epsilon}

\def\bbE{\mathbb{E}}

\def\bbP{{\mathbb{P}}}

\def\bbZ{\mathbb{Z}}

\newcommand{\bra}[1]{\langle #1 |}
\newcommand{\ket}[1]{| #1 \rangle}

\newcommand{\proj}[1]{| #1 \>\!\< #1 |}

\DeclareMathOperator{\rank}{rank}

\DeclareMathOperator{\tr}{Tr}

\def\*{\star}

\def\tilde{\widetilde}

		 \def\cB{{\cal B}}		 
\def\cC{{\cal C}}
\def\cD{{\cal D}}		 \def\cE{{\cal E}}

		 \def\cN{{\cal N}}

\def\cX{{\cal X}}
\def\cY{{\cal Y}}		 

\begin{document}

\title{Quantum error correcting codes based on privacy amplification}
\date{\today}
\author{Zhicheng Luo}
\email{zluo@usc.edu}
\affiliation{Physics Department, University of Southern California, Los Angeles, CA 90089, USA}

\begin{abstract}
Calderbank-Shor-Steane (CSS) quantum error-correcting codes are 
based on pairs of classical codes which are mutually dual containing. 
Explicit constructions of such codes for large blocklengths and
with good error correcting properties are not easy to find.
In this paper we propose a construction of CSS codes which combines a classical code
with a two-universal hash function. We show, using the results of Renner and Koenig~\cite{RK04},
that the communication rates of such codes approach the hashing bound on tensor powers of
Pauli channels in the limit of large block-length. 
While the bit-flip errors can be decoded as efficiently as the classical code used, 
the problem of efficiently decoding  the phase-flip errors remains open.

\end{abstract}

\maketitle

\section{Introduction}
Secret classical information and quantum information are intimately 
related~\cite{Sch98, CP2002}. One can convert a maximally entangled state 
shared by two distant parties into a secret classical key by local bilateral measurements in the standard
basis. Since it is a pure state, it is decoupled from the environment, and so is the information about the measurement 
outcome. An important application of this simple observation is the Shor-Preskill~\cite{SP00} proof of the 
security of the Bennett-Brassard 1984 (BB84)~\cite{BB84} quantum key
distribution (QKD) protocol. They convert an entanglement distillation protocol based on
Calderbank-Shor-Steane (CSS) quantum error correcting codes~\cite{CS96, Steane96} into a key distribution protocol. 
The correspondence often also works in the other direction. Given a quantum channel or noisy bipartite quantum state,
a large class of secret communication protocols can be made into quantum communication protocols 
by performing all the steps ``coherently'', i.e. replacing probabilistic mixtures by quantum superpositions
~\cite{Devetak03, DW03b, DW03c}. This result was derived in the idealized asymptotic context of quantum Shannon 
theory, without giving explicit, let alone efficient, code constructions. Not coincidentally, these asymptotic codes
 had a structure reminiscent of CSS codes.

In this paper we bridge the gap between these asymptotic constructions and CSS codes. We obtain a
subclass of CSS codes, called P-CSS codes, by making coherent a class of private codes 
studied by Renner and Koenig~\cite{RK04}. The original CSS ~\cite{CS96, Steane96} construction 
requires  two classical linear finite distance codes which contain each other's duals.
One is used for correcting bit flip errors, while the other corrects phase flip errors and 
was identified in ~\cite{SP00} as being responsible for privacy amplification.
The resulting quantum code is also of finite distance. 

P-CSS codes involve a combination of a classical error correcting code 
and a privacy amplification protocol.
Consequently, the dual-containing property does not come into play,
which greatly simplifies the construction. On the other hand, P-CSS codes are not finite distance codes.
This is not necessarily a problem: modern coding theory (eg. LDPC~\cite{LDPC} and turbo codes~\cite{BGT93}) 
focuses not on distance
but on performance on simple i.i.d. (independent, identically distributed) channels. 
The main result of this paper is to show that P-CSS 
codes  have excellent asymptotic behaviour, attaining the hashing bound on i.i.d. Pauli channels.

The paper is organized as follows. In section II, we recall the definition of two-universal 
hash functions and  how they are used in private codes. We then explain how they can be turned into
quantum codes, and give examples.
In section III, we state and prove the main theorem which quantifies the performance of our 
code in terms of the parameters of the underlying classical error correcting code and privacy amplification
protocol. The asymptotic rates for entanglement transmission  are calculated in section IV for memoryless 
qubit Pauli channels. We conclude in section V with open problems.   

\section{Quantum codes based on affine two-universal hash functions}

\subsection{Private classical codes by two-universal hashing}

Assume a communication scenario in which a sender Alice is connected to 
a receiver Bob and eavesdropper Eve through a noisy classical ``wiretap'' channel with one input
and two outputs~\cite{Wyner75Bell}. Alice wants to send  messages to Bob about 
which Eve is supposed to find out as little as possible. There are two obstacles 
Alice and Bob need to overcome: i) Bob receives only a noisy copy of Alice's input, 
and ii)  (partial) information about her input leaks to Eve.
In order to reduce Bob's noise they must apply \emph{error correction}, and in order to
increase Eve's noise they must perform \emph{privacy amplification}. Together, the two form
a \emph{private code}.

An $[n, k]$ linear error correcting code $C$ is a $k$-dimensional linear subspace of $\bbZ_2^n$. 
It can be given either as the column space of an ${n\times k}$ \emph{generator matrix} $G$,
so that $C = \{G y: y \in \bbZ_2^k \}$ \footnote{all vectors are assumed column vectors},
or the null space of an ${(n - k)\times n}$ \emph{parity check matrix} $H$.
The row space of  $H$ is the dual code $C^{\perp}$ of $C$. 
The interpretation is that Alice encodes her $k$-bit message $y$ into the $n$-bit codeword
$x = G y$. This is sent down a noisy $n$-bit channel to Bob, who then tries to decode
the original message $y$. 

Privacy amplification (PA) protocols~\cite{BBR88, ILL89, BBCM95, RK04} are defined in terms of random functions.
A random function from $\cX$ to $\cY$ is a random variable taking values from the set of functions with
domain $\cX$ and range $\cY$. A random function $f$ from $\cX$ to $\cY$ is called \emph{two-universal}
if
$$
\Pr{[f(x) = f(x')]} \leq \frac{1}{|\cY|}~,
$$
for any distinct $x, x'\in \cX$. 

Suppose Alice has a $k$ bit wiretap channel which is noiseless to Bob but partially noisy to Eve.
Still Eve receives some information about the identity of the bits, and 
the $k$ bits are thus not secure. 
However, Alice will settle for transmitting a smaller number of bits $m$, if it ensures
that they will now be secret from Eve. The PA protocol is characterized by  a two-universal random
function $f: \bbZ_2^k \rightarrow  \bbZ_2^m$.
Alice starts by  drawing a particular realization of $f$.
To convey the secret message $s \in \bbZ_2^m$
she encodes it as an equiprobably chosen random element $y$ of the set 
$Z_s(f) = \{y: f(y) = s \}$, and sends that  down the channel. She then publicly
announces the realization of $f$.
Now Bob, knowing $y$ and $f$, also knows $s$. On the other hand, it can be shown~\cite{BBR88, ILL89, BBCM95} 
that Eve's correlation with $s$ is exponentially small in $k - m$.
 
A private code is a combination of an error correcting code $C$ and a privacy amplification
protocol $f$. Formally it is a set of sets $\{C_s(f): {s \in \bbZ_2^m} \}$, where 
$C_s(f) = \{x: x=G y, f(y)=s\}$. It is used for transmitting $m$ bits of 
approximately secret information reliably over an $n$-bit wiretap channel which is noisy to Bob.
As in the noiseless case, Alice draws a particular realization of $f$, and 
encodes the secret message $s \in \bbZ_2^m$ as an equiprobably chosen random element of the set $C_s(f)$.
After transmission she publicly announces the realization of $f$ to enable Bob to decode the message.


\subsection{The P-CSS code construction}

Following \cite{Devetak03}, we can construct  quantum codes by coherifying 
the private code $\{C_s(f): {s\in\bbZ_2^m}\}$. We will work in the computational
qubit orthonormal basis $\{ \ket{0}, \ket{1} \}$ and the corresponding 
$n$-qubit basis $\{ \ket{x}: x \in  \bbZ_2^n \}$, where 
$\ket{(x_1,x_2, \dots x_n)} = \ket{x_1} \ket{x_2} \dots  \ket{x_n}$.
An $[[n, k]]$ quantum code ${\cal C}$ is a $2^k$-dimensional subspace of the space of $n$-qubits
(see eg. \cite{NC00} and references therein).
It is used to encode $k$ qubits into $n$, by unitarily transforming the standard $k$-qubit basis
vectors into a particular basis of $\cal C$. 
The basis vectors of $\cal C$ are referred to as the quantum codewords of $\cal C$.
Given the private code  $\{C_s(f): {s\in\bbZ_2^m}\}$, we define its quantum counterpart $\cal C$
by quantum codewords $\{\ket{\varphi_s(f)}: {s\in\bbZ_2^m}\}$, with 
\be
\ket{\varphi_s(f)} = \frac{1}{\sqrt{2^{k - m}}} \sum_{x\in C_s(f)} \ket{x}.
\label{cword}
\ee
More precisely, this defines a random quantum code, as for each realization of $f$ we 
have a potentially different $\cal C$. We may occasionally be sloppy about this distinction.

Recall that a quantum code is called a stabilizer code if it is the simultaneous
$+1$ eigenspace of a set of $k$ independent $n$-qubit Pauli operators (called the ``stabilizers''). 
A CSS code is a stabilizer code, each stabilizer of which is composed of purely $Z$ or purely $X$ operators.
It is defined by two classical error correcting codes $C_1$ and $C_2$, with parity check matrices
$H_1$ and $H_2$, respectively, such that $C_2^{\perp} \subseteq C_1$.
 The set of stabilizers of the CSS code is given by $\{Z^{H_1}, X^{H_2}\}$, where
$Z^H = \{Z^h : \, h \text{\,\, is a row of}\, H\}$, etc., and 
$Z^h = Z^{h_1} \otimes \dots Z^{h_n}$, $h = (h_1, \dots, h_n)$. 
Bit flip errors are corrected by the $Z$ stabilizers, and phase flip errors
are corrected by the $X$ stabilizers. CSS codes can alternatively be given in
terms of codewords:
\be
\ket{x + C_2^\perp} = \frac{1}{\sqrt{|C_2^\perp|}} \sum_{y \in C_2^\perp} \ket{x + y}
\label{csscw}
\ee
where $x$ runs over the elements of $C_1$. It is easy to see that $\ket{x + C_2^\perp}$
only depends on the coset of $C_1/C_2^{\perp}$ to which $x$ belongs, so we can
let $x$ only run over suitably chosen coset representatives $x_s$.

The expression (\ref{csscw}) very much resembles (\ref{cword}). 
We next show that if the random function $f$ is affine, or rather that
each of its realizations is affine, then each realization of $\cal C$ is indeed a CSS code.
Assume $f$ is of the form
\be
s = f(y) =  A y + s_0,
\label{affine}
\ee
for some $s_0 \in \bbZ_2^m$ and  full rank $m \times k$ matrix $A$.
Denoting the null space of $A$ by $C''$, we see that $y$ lies in the coset of 
$\bbZ_2^k/ C''$  determined by $s - s_0$. 
Let $F$ be a $k \times (k - m)$ matrix whose column space is equal to $C''$ 
(if we think of $A$ as a sort of parity check matrix, then $F$ would be the 
corresponding generator matrix). Then the set $Z_s(f) = \{y: f(y) = s \}$ can be written as 
$$
Z_s(f) = \{ F t + y_s: t \in \bbZ_2^{k - m}\},
$$
where the $y_s$ are the coset representatives of $\bbZ_2^k/ C''$.
Hence $C_s(f) = \{x: x=G y, y \in Z_s(f)\}$ can be written as
$$
C_s(f) = \{ G F t + x_s: t \in \bbZ_2^{ k - m}\},
$$
where $x_s = G y_s$ are the coset representatives of $C/C'$,
and $C'$ is the column space of the $n \times (k - m)$ matrix $G F$.
Thus the quantum code $\cal C$ with codewords (\ref{cword}) is 
the CSS code composed from the classical codes $C$ and $(C')^\perp$
with parity check matrices $H$ and $(G F)^T$, respectively.
Its stabilizers are thus $\{Z^{H}, X^{(G F)^T)}\}$. Observe that the quantum
code $\cal C$ is independent of the constant term $s_0$ from the definition of $f$ (\ref{affine}); 
its only effect is to permute the quantum codewords.

To recapitulate, 
$H$ and $G$ are the parity check matrix and generator matrix of the classical
code $C$, respectively. $F$ comes from the affine structure of the two universal hash function $f$.
Since $H G = 0$, we also have $H (GF) = 0$, so the bit flip and phase flip codes
are automatically  contained in each other's duals. We dubbed the CSS codes 
thus obtained \emph{P-CSS codes}.

\subsection{Examples}
We will illustrate our construction by using
a particular class of affine two-universal hash functions 
taken from ~\cite{BBR88, WC81}. 
There is a natural bijection between $\bbZ_2^k$ and polynomials in 
$\zeta\in GF(2^k)$ ($\zeta^{2^k-1} = 1$) of degree $k-1$ over $GF(2)$ : 
$$
y = (y_0, ... , y_{k-1}) \,\Leftrightarrow\,  y(\zeta) = \sum_{i=0}^{k-1} y_i\zeta^i~.
$$
Therefore, we can define linear maps :
$$ \alpha(y(\zeta)) = y ~~~\text{and}~~~ \alpha^{-1}(y) = y(\zeta)~.$$
For $a,b\in GF(2^k)$ ($a\neq 0$), the polynomial 
$$q_{a,b}(y(\zeta)) = a y(\zeta)  + b$$ 
defines a permutation of $GF(2^k)$, which induces 
a permutation $\pi_{a,b} : \bbZ_2^k \ra \bbZ_2^k$ by 
$$\pi_{a,b} = \alpha \circ q_{a,b} \circ \alpha^{-1}.$$ 
For any fixed $m\leq k$, define the  map $\tau: \bbZ_2^k \ra \bbZ_2^m$
which projects onto the first $m$ bits:
$$
\tau((y_1, ... , y_k)) = (y_1, ... , y_m).
$$
Define the function $h_{a,b}(y) : \bbZ_2^k \ra \bbZ_2^m$ as
the composition 
$$h_{a,b} = \tau \circ \pi_{a,b},$$
and the set 
$$
\bbP = \{h_{a,b}|a,b\in GF(2^k), a\neq 0\}.
$$
The random function $f$ which is uniformly distributed on
$\bbP$ is known to be two-universal.
Moreover, each $h_{a,b}$ is a composition of affine functions, and
hence affine. As observed earlier, the resulting CSS code
will be independent of the value of $b$.

Our first set of examples are $[[7 ,1 ]]$ CSS codes. We take $n = 7$, 
$k = 4$ and $m = 1$. For $C$ we take the $[7,4,3]$ Hamming code
given by 
\begin{align*}
G^T = \left(
  \begin{aligned}
     & 1 \,\, 0 \,\, 0 \,\, 0 \,\, 0 \,\, 1 \,\, 1 \\
     & 0 \,\, 1 \,\, 0 \,\, 0 \,\, 1 \,\, 0 \,\, 1 \\
     & 0 \,\, 0 \,\, 1 \,\, 0 \,\, 1 \,\, 1 \,\, 0 \\
     & 0 \,\, 0 \,\, 0 \,\, 1 \,\, 1 \,\, 1 \,\, 1
  \end{aligned}\right)~,\,\,\,
H = \left(
  \begin{aligned}
     & 0 \,\, 0 \,\, 0 \,\, 1 \,\, 1 \,\, 1 \,\, 1 \\
     & 0 \,\, 1 \,\, 1 \,\, 0 \,\, 0 \,\, 1 \,\, 1 \\
     & 1 \,\, 0 \,\, 1 \,\, 0 \,\, 1 \,\, 0 \,\, 1 
  \end{aligned}\right)~,  
\end{align*}

Different values of $a\in GF(16)$ will in general give rise to different CSS codes. 
If we choose $a = \zeta^{-2}$ ($\zeta^{15}=1$), then 
\begin{align*}
F = \left(
  \begin{aligned}
     & 0 \,\, 1 \,\, 0 \\
     & 0 \,\, 1 \,\, 1 \\
     & 0 \,\, 0 \,\, 1 \\
     & 1 \,\, 0 \,\, 0
  \end{aligned}\right)~,\,\,\,  
(GF)^T = \left(
  \begin{aligned}
     & 0 \,\, 0 \,\, 0 \,\, 1 \,\, 1 \,\, 1 \,\, 1 \\
     & 1 \,\, 1 \,\, 0 \,\, 0 \,\, 1 \,\, 1 \,\, 0 \\
     & 0 \,\, 1 \,\, 1 \,\, 0 \,\, 0 \,\, 1 \,\, 1
  \end{aligned}\right)~.  
\end{align*}
The constructed seven qubit quantum code is none other than the $[[7, 1, 3]]$
Steane code with stabilizers $\{Z^H, X^{(GF)^T}\} = $ 
\begin{align*}
\left\{
   \begin{aligned}
   & Z_4Z_5Z_6Z_7, \,\,\,\,Z_2Z_3Z_6Z_7, \,\,\,\,\,Z_1Z_2Z_5Z_6, \\
   & X_4X_5X_6X_7, \,X_2X_3X_6X_7, \,X_1X_2X_5X_6
   \end{aligned}\right\}~.
\end{align*}
On the other hand, if we choose $a=\zeta$, we have 
\begin{align*}
F = \left(
  \begin{aligned}
     & 1 \,\, 0 \,\, 0 \\
     & 0 \,\, 1 \,\, 0 \\
     & 0 \,\, 0 \,\, 1 \\
     & 0 \,\, 0 \,\, 0
  \end{aligned}\right)~,\,\,\,
(GF)^T = \left(
  \begin{aligned}
     & 1 \,\, 0 \,\, 0 \,\, 0 \,\, 0 \,\, 1 \,\, 1 \\
     & 0 \,\, 1 \,\, 0 \,\, 0 \,\, 1 \,\, 0 \,\, 1 \\
     & 0 \,\, 0 \,\, 1 \,\, 0 \,\, 1 \,\, 1 \,\, 0
  \end{aligned}\right)~.  
\end{align*}
This code has stabilizers 
\begin{align*}
\left\{
   \begin{aligned}
   & Z_4Z_5Z_6Z_7, \,Z_2Z_3Z_6Z_7, \,Z_1Z_2Z_5Z_6, \\
   & \,\,X_1X_6X_7, \,\,\,\,\,X_2X_5X_7, \,\,\,\,\,X_3X_5X_6
   \end{aligned}\right\}~.
\end{align*}
which can not correct the $Z_4$ error, and hence does not have distance $3$.

\section{Entanglement transmission over Pauli channels}

From the examples from the previous section we saw that, unlike the original
CSS construction, the P-CSS construction does not tell us anything about 
the distance of the obtained quantum code. So in what sense is this construction useful?
What can we say about the error correction properties of P-CSS codes?

What can be quantified is the performance of P-CSS codes on Pauli channels.
In this section we will define the communication scenario of interest, 
and prove a theorem about the quality of entanglement that can be transmitted
through such channels, given the properties of the classical code $C$ and 
parameters of the hash function $f$ which comprise the P-CSS code.
The central tool is a theorem by Renner and Koenig~\cite{RK04} regarding quantum privacy amplification
(see Appendix).

An $n$-qubit Pauli channel ${\cN_n}$, which applies Pauli errors $X^{u}Z^{v}$ with probability 
$p_{u,v}$, can be described as 
$$
{\cN_n}(\rho) = \sum_{u,v \in \bbZ_2^n} p_{u,v} X^{u}Z^{v} \rho Z^{v}X^{u}~,
$$
where $u = (u_1,...,u_n)$ and $v = (v_1,...,v_n)$ are binary vectors of length $n$.
We can equivalently represent  the action of ${\cN_n}$ as
$$
{\cN_n}(\rho) = \tr_E U^{A \ra BE}_{\cN_n} (\rho),
$$
where the isometry $U_{\cN_n}$ is defined by its action on the computational 
basis states $\ket{x}^{A}, x \in \bbZ_2^n$, as
\be
U_{{\cN_n}}\ket{x}^{A} = \ket{\phi_x}^{BE} = \sum_{u,v}\sqrt{p_{u,v}}(-1)^{x\cdot v}\ket{x+u}^{B}\ket{u,v}^{E}.
\label{standard}
\ee 
The interpretation is that, in being sent through the channel, Alice's pure input state $\ket{x}^{A}$
undergoes an isometry which splits it up between Bob's output system $B$ and the unobservable environment $E$.
This is analogous to the wire-tap channel with the environment playing the role of the eavesdropper Eve.
Both Eve and Bob receive the state only partially.

We are concerned with the task of entanglement transmission.
Alice is handed the $A'$ part of the $m$ ebit state $(\ket{\Phi}^{\otimes m})^{RA'}$,
where
$$
\ket{\Phi} = \frac{1}{\sqrt{2}} (\ket{0}\ket{0} + \ket{1}\ket{1}),
$$ 
the other half of which belongs to a reference system $R$ which Alice cannot access.
The objective is to use the channel ${\cN_n}$ to transfer the purification of $R$ to Bob, so that Bob ends up
sharing an $m$ ebit state $(\ket{\Phi}^{\otimes m})^{RB'}$ with $R$. In most cases this can be done only approximately.

\medskip

Formally, an $[[n, m]]$ entanglement generation code consists of an encoding
isometry $U_{\cC}^{A' \rightarrow A}$ (which takes the standard $m$-qubit basis
to the codewords of $\cC$) and a decoding operation $\cD^{B \rightarrow B'}$.
It is said to be $\eta$-good if
$$
\left\|\cD \circ {\cN_n} \circ U_{\cC} (\Phi^{\otimes m})^{RA'} 
-  (\Phi^{\otimes m})^{RB'} \right\|_1  \leq \eta.
$$
 
The P-CSS code $\cC$ which we will use for entanglement transmission
consists of an $[n, k]$ code $C$ with parity check matrix $H$, and
an affine random two-universal hash function $f: \bbZ_2^k \rightarrow \bbZ_2^m$.
The performance of the quantum code will be given in terms of the performance
of the classical code $C$, which involves the bit flip errors $u$. 
To decode $C$ one measures the error syndrome $e = H u$. While the syndrome is 
uniquely determined by the error, the reverse is usually not the case.
One needs to define the ``inverse syndrome'' function 
$\hat{u}:  \bbZ_2^{n - k} \rightarrow \bbZ_2^n $, for which  $H \hat{u}(e) = e$.
Now $G$, $H$ and $\hat{u}$ fully specify the classical encoding and decoding operations.
Define
\begin{eqnarray}
p_u & = & \sum_v p_{u,v} \\
p_e & = & \sum_{u : H u~=~e}p_{u} \label{parameter1} \\
p_{u,v|e} & = & I(H u~=~e)  p_{u,v}/p_e, \\
p_{u|e} & = & I(H u~=~e) p_u/p_e, \label{parameter2} \\
\eps(e) & = & 1 - p_{\hat{u}(e) | e},\\
\eps & = & \sum_e p_e \eps(e). \label{parameter3}
\end{eqnarray}
Here $p_u$ is the marginal probability of bit flip error $u$ occurring,
$p_e$ is the probability of the bit flip error syndrome being $e$,
$p_{u|e}$ is the probability of the bit flip error being $u$ conditional
on observing the syndrome $e$, and $\eps(e)$ and $\eps$ are the probabilities of 
incorrectly identifying the bit flip error with and without conditioning on the syndrome $e$,
respectively.

We can now formulate our main theorem.

\begin{thm}
Given an $n$-qubit Pauli channel ${\cN_n}$, a classical linear $[n, k]$ code $C$ with average error probability
$\epsilon$, and an affine random two-universal hash function $f: \bbZ_2^k \rightarrow \bbZ_2^m$,
there exists a random $[[n, m]]$ P-CSS entanglement generation code
which is on average $\eta$-good on ${\cN_n}$, with
$$
\eta = 2\sqrt{2\eps' + 4\sqrt{2\eps}} + 2\sqrt{2\eps}~,
$$  
$\eps'= 2^{-\frac{1}{2}(H_2(XE)_{\omega} - H_0(E)_{\omega} + k - n - m)}$, and 
\be
\omega^{XBE} = \frac{1}{2^n}\sum_{x\in \bbZ_2^n}\ket{x}\bra{x}^X\otimes \ket{\phi_x}\bra{\phi_x}^{BE}~.
\label{omegaXBE}
\ee 
$H_2$ and $H_0$ denote R\'enyi entropies (see Appendix for details), and the state $\omega^{XBE}$ 
is the result of sending a randomly chosen computational
basis element through the Pauli channel ${\cN_n}$.
\end{thm}

\begin{proof}
We will break up the proof into a number of steps.

1. Imagine Alice sends  the state $\ket{x}$, ${x\in C}$, through the channel, resulting in
the state $\ket{\phi_x}^{BE}$ defined in (\ref{standard}).
Measuring the stabilizers $\{Z^H\}$, with probability $p_e$, Bob will 
get the syndrome $e$ and state 
$$
\ket{\phi_x(e)}^{BE} = \sum_{u : H u = e,v}\sqrt{p_{u,v|e}}(-1)^{x\cdot v}\ket{x + u}^{B}\ket{u,v}^{E},
$$
where $p_e$ and $p_{u,v|e}$ are defined in ($\ref{parameter1}$) and ($\ref{parameter2}$).
Bob applies $X^{\hat{u}(e)}$ to correct the bit flip errors, resulting in  
\begin{align*}
\ket{\phi'_x(e)}=\sqrt{1-\eps(e)} \ket{x}^B \ket{\psi^{good}_{x}(e)}^{E}+\sqrt{\eps(e)}\ket{\psi^{bad}_{x}(e)}^{BE},
\end{align*}
where 
$$
\ket{\psi^{good}_{x}(e)}^{E} = \ket{\hat{u}(e)}^{E_1}
\sum_v \sqrt{p_{v|\hat{u}(e)}} (-1)^{x\cdot v} \ket{v}^{E_2},~
$$ 
and $\ket{\psi^{bad}_{x}(e)}^{BE}$ corresponds to $\{u\neq \hat{u}(e): Hu=e\}$,
i.e. the errors that fail to get corrected.

Since $\ket{\psi^{bad}_{x}(e)}^{BE}$ is orthogonal to $\ket{x}^B\ket{\psi^{good}_{x}(e)}^{E}$,
we have (see Appendix for details about the fidelity $F$)
\be
F\left(\ket{\phi'_x(e)}^{BE}, \ket{x}^{B}\ket{\psi^{good}_{x}(e)}^{E} \right) = 1 - \eps(e)~.
\label{fidelity}
\ee

2. If Alice send the quantum codeword 
$$
\ket{\varphi_s} = \frac{1}{\sqrt{2^{k - m}}}\sum_{z \in C'} \ket{x_s + z},
$$
and Bob performs the same steps as above, the resulting state is 
\be
\frac{1}{\sqrt{2^{k - m}}} \sum_{z \in C'}\ket{\phi'_{x_s + z}(e)}^{BE},
\label{state1}
\ee
which  has fidelity $1 - \eps(e)$ with 
\be
\frac{1}{\sqrt{2^{k - m}}}\sum_{ z\in C'}  \ket{x_s + z}^{B}\ket{\psi^{good}_{x_s + z}(e)}^{E}.
\label{state2}
\ee

3. Let $\{ \cE_i = v_i + C'^{\bot}: i\in\bbZ_2^{k-m} \}$
denote the cosets of $C'^{\bot}$ in $\bbZ^n_2$, with $C'$ defined
as in Section IIB. Recall also that $\{ x_s: s \in \bbZ_2^{m} \}$ were defined as the 
coset representatives of $C/C'$.
 Since $z\cdot v = z\cdot v_i$ for all $v\in\cE_i$ and $z\in C'$, 
we can rewrite the state (\ref{state2}) as
\be
\sum_{ i\in\bbZ_2^{k-m}} \ket{\varphi_{s,i}}^{B}\ket{{\theta}_{s,i}(e)}^{E}~,
\label{state3}
\ee
with
$$
\ket{\varphi_{s,i}}^{B} = \frac{1}{\sqrt{2^{k-m}}}\sum_{z\in C'}(-1)^{z\cdot v_i}\ket{z+x_s}^{B} ~,
$$
\be
\ket{{\theta}_{s,i}(e)}^{E} = \ket{\hat{u}(e)}^{E_1}\otimes\sum_{v\in\cE_i} \sqrt{p_{v|\hat{u}(e)}} 
(-1)^{x_s\cdot v}\ket{v}^{E_2}~.
\label{thetasiE}
\ee
Observe that $\{ \ket{\varphi_{s,i}}^B: s \in \bbZ_2^{m}, i\in\bbZ_2^{k-m} \}$ are
a basis for the system $B$, and there exists a Clifford unitary 
$U: \ket{\varphi_{s,i}}^B \mapsto \ket{s}^{B_1}\ket{i}^{B_2}$.


Bob applies $U$, resulting in  a state $\ket{\Upsilon_s(e)}^{B_1B_2E}$
which has fidelity $1 - \eps(e)$ with
\be
\ket{\tilde{\Upsilon}_s(e)}^{B_1B_2E} = \ket{s}^{B_1} \sum_i \ket{i}^{B_2}\ket{\theta_{s,i}(e)}^E.
\label{tupsilon}
\ee 

4. Now we need to ensure that there is no $s$ dependence of the state of  the environment $E$. 
As we will see this has to do with the performance of the private 
code on which our P-CSS code is based. 
 
By Lemma 1 below,
\be
\frac{1}{2^m}\sum_s \bbE_f\left\|\sigma_s^E - \sigma_0^E\right\|_1 \leq 2\eps',
\label{PA2}
\ee 
where $\sigma^E_s=\frac{1}{2^{k-m}}\sum_{z\in C'}\phi_{z+x_s}^E$, $\phi_{x}^E = \tr_B \ket{\phi_{x}}\bra{\phi_{x}}^{BE}$, 
and  $\eps' = 2^{-\frac{1}{2}(H_2(XE)_{\omega} - H_0(E)_{\omega} + k - n - m)}$.
It is easy to see that
$$
\tilde{\Upsilon}_s(e)^E = \frac{1}{2^{k-m}}\sum_{z\in C'}\psi^{good}_{x_s + z}(e)^E.
$$
On the other hand, since $\phi_{x_s + z}^E = \sum_e p_e \phi_{x_s + z}(e)^E$, it follows that
$$
\sigma^E_s =  \sum_e p_e \frac{1}{2^{k-m}}\sum_{z\in C'}\phi_{x_s + z}(e)^E.
$$
From  the concavity ($\ref{concavity}$) and monotonicity ($\ref{monotonicity}$) of fidelity, and (\ref{fidelity})
we have 
\begin{align*}
\begin{aligned}
 & F\left(\sigma_s^{E},\sum_e p_e \tilde{\Upsilon}_s(e)^E \right)\\
 &\geq \left(\sum_{e,z} \frac{p_e}{2^{k-m}}
 \sqrt{F\left(  \phi_{x_s + z}(e)^E , \psi^{good}_{x_s + z}(e)^E  \right)}\right)^2 \\
        &\geq \left(\sum_e p_e(1 - \eps(e))\right)^2 \geq 1 - 2\eps~.
\end{aligned}
\end{align*}
Then by $(\ref{tracefidelity})$, for all $s$ we have 
\be
\left\|\sigma_s^{E} - \sum_e p_e \tilde{\Upsilon}_s(e)^E \right\|_1 \leq 2\sqrt{2\eps}~.
\label{tracedistance}
\ee
By the triangle inequality for trace distance, we can combine ($\ref{PA2}$) and ($\ref{tracedistance}$) to get
\be
\frac{1}{2^m}\sum_s \bbE_f\left\|\sum_e p_e \tilde{\Upsilon}_s(e)^E - \sum_e p_e \tilde{\Upsilon}_0(e)^E    \right\|_1 \leq 2\eps' + 4\sqrt{2\eps}~.
\label{PA3}
\ee
Since $\tilde{\Upsilon}_s(e)^E = \sum_{i}\ket{\theta_{s,i}(e)}\bra{\theta_{s,i}(e)}^E$ and
$\{\theta_{s,i}(e)\}_{i,e}$  is an orthogonal set,
we can express the privacy condition ($\ref{PA3}$) as  
$$
\frac{1}{2^m}\sum_{s,e,i} p_e\bbE_f\left\|\theta_{s,i}(e)^E - \theta_{0,i}(e)^E\right\|_1 \leq 2\eps' + 4\sqrt{2\eps}~.
$$
Set $\ket{\theta_{s,i}(e)}^E = \sqrt{q_i(e)}~\ket{\hat{\theta}_{s,i}(e)}^E$, 
where $\ket{\hat{\theta}_{s,i}(e)}^E$ is normalized, 
$q_i(e)=\sum_{v\in\cE_i}p_{v|\hat{u}(e)}$ depends on $f$ through $\cE_i$, 
and $\sum_iq_i(e)=1$. Thus by $(\ref{tracefidelity})$ we have
\be
\frac{1}{2^m}\sum_{s,e,i} p_e \bbE_f \left[q_i(e)\left(1-\left|\langle \hat{\theta}_{s,i}(e)|\hat{\theta}_{0,i}(e)\rangle\right|\right)\right] \leq \eps' + 2\sqrt{2\eps}~.
\label{PA4}
\ee

5. Steps 2-4 considered Alice sending a single codeword through the channel.
Now we are ready for the actual task of entanglement transmission.
Alice is handed the $A'$ part of the state 
$$
 (\Phi^{\otimes m})^{RA'} =  \frac{1}{\sqrt{2^{m}}} \sum_{s\in \bbZ_2^m} \ket{s}^R \ket{s}^{A'}.
$$
She performs the encoding $U_\cC: \ket{s}^{A'} \mapsto \ket{\varphi_s}^{A'}$
and sends the output down the channel. Bob performs all the operations in
steps 2-4 and the resulting state (conditional on syndrome $e$) is
$$
\ket{\Upsilon(e)} =  \frac{1}{\sqrt{2^{m}}} \sum_s \ket{s}^R \ket{{\Upsilon}_s(e)}^{B_1B_2E}.
$$
Define also
$$
\ket{\tilde{\Upsilon}(e)} =  \frac{1}{\sqrt{2^{m}}}\sum_s \ket{s}^R \ket{\tilde{\Upsilon}_s(e)}^{B_1B_2E}, 
$$
and
$$
\ket{\hat{\Upsilon}(e)} = \frac{1}{\sqrt{2^m}}\sum_{s,i}\ket{s}^{R}\ket{s}^{B_1}(-1)^{b(s,i)} 
\ket{i}^{B_2} \ket{\theta_{0,i}(e)}^E,
$$
with
$$
(-1)^{b(s,i)} = \frac{\langle \hat{\theta}_{s,i}(e)|\hat{\theta}_{0,i}(e)\rangle}{\left|\langle \hat{\theta}_{s,i}(e)|\hat{\theta}_{0,i}(e)\rangle\right|}~.
$$
Thus
\begin{align*}
\langle \tilde{\Upsilon}(e)|\hat{\Upsilon}(e) \rangle &= \frac{1}{2^m}\sum_{s,i}q_i(e) (-1)^{b(s,i)}\langle \hat{\theta}_{s,i}(e)|\hat{\theta}_{0,i}(e)\rangle \\
&=  \frac{1}{2^m}\sum_{s,i}q_i(e)\left|\langle \hat{\theta}_{s,i}(e)|\hat{\theta}_{0,i}(e)\rangle\right|~.
\end{align*}
Averaging over all the possible syndromes, we can define 
$$
\Upsilon = \sum_{e} {p_e} \Upsilon(e)^{RB_1B_2E}~,
$$
and similarly $\tilde{\Upsilon}$ and $\tilde{\Upsilon}_0$.

Then by the concavity ($\ref{concavity}$) of fidelity, the property of convex function (i.e. $\bbE X^2 \geq (\bbE X)^2$) and the privacy condition ($\ref{PA4}$), we have
\begin{align*}
&~~~~~\bbE_f F(\hat{\Upsilon}^{RB_1B_2E}, \tilde{\Upsilon}^{RB_1B_2E})\\ 
&\geq \bbE_f\left(\sum_{e} p_e \sqrt{F\left(\ket{\hat{\Upsilon}(e)}, \ket{\tilde{\Upsilon}(e)}\right)}\right)^2\\
&\geq \left(\frac{1}{2^m}\sum_{s,i,e}p_e\bbE_f\left[q_i(e)\left|\langle \hat{\theta}_{s,i}(e)|\hat{\theta}_{0,i}(e)\rangle\right|\right]\right)^2\\
&\geq 1 - 2\eps' - 4\sqrt{2\eps}~.
\end{align*}
It is not hard to see that 
$$
F\left(\Upsilon^{RB_1B_2E}, \tilde{\Upsilon}^{RB_1B_2E}\right) \geq 1 - 2\eps~,
$$
since by the concavity ($\ref{concavity}$) of fidelity, the fact that trace-preserving quantum 
operations never reduce fidelity, 
and condition ($\ref{fidelity}$) we have 
\begin{align*}
\begin{aligned}
F\left(\Upsilon, \tilde{\Upsilon} \right) &\geq \left(\sum_{e} p_e 
\sqrt{F\left(\Upsilon(e), \tilde{\Upsilon}(e)\right)}\right)^2 \\
        &\geq \left(\sum_e p_e(1 - \eps(e))\right)^2 \geq 1 - 2\eps~.
\end{aligned}
\end{align*}
Combining the results above, by ($\ref{tracefidelity}$) we have
\begin{align*}
\bbE_f\|\Upsilon - \hat{\Upsilon}\|_1 &\leq \bbE_f\|\Upsilon - \tilde{\Upsilon}\|_1 + \bbE_f\|\tilde{\Upsilon} -
 \hat{\Upsilon}\|_1 \leq \eta 
\end{align*}
with $\eta = 2\sqrt{2\eps' + 4\sqrt{2\eps}} + 2\sqrt{2\eps}$.

Finally Bob performs the decoupling unitary
$$V^{B_1B_2} = \sum_{s,i}(-1)^{b(s,i)}\ket{s}\bra{s}^{B_1}\otimes\ket{i}\bra{i}^{B_2},$$
and throws away the $B_2$ system (and also implicitly $E$ which he never had access to anyway).
This combined operation takes $\hat{\Upsilon}$ to the desired state
\be
 \frac{1}{\sqrt{2^{m}}} \sum_{s\in \bbZ_2^m} \ket{s}^R \ket{s}^{B_1}.
\label{desire}
\ee
By the monotonicity of trace distance ($\ref{monodis}$),
it takes the actual state $\Upsilon$ to a state which, on average, is $\eta$-close to 
(\ref{desire}) in trace distance. Hence the average performance of the code (averaging over $f$)
is $\eta$-good, as claimed. This averaging can be treated in two ways. Alice and Bob could
start with pre-shared randomness, based on which they choose $f$. More simply, if the 
codes are good on average, then at least one does at least as well as the average. 

It is worth summing up Bob's decoding operation.
He first measures the stabilizers $\{Z^H\}$, obtaining the bit flip error
syndrome $e$, and applies $X^{\hat{u}(e)}$ to correct the bit flip errors.
To deal with the phase errors, he performs $U^{B \ra B_1 B_2}$ followed by $V^{B_1 B_2}$ and discarding $B_2$.
This phase-correcting  combined operation can be implemented in a simpler, less coherent way.
Instead of performing $U$, he measures the stabilizers $\{X^{(GF)^T}\}$, 
thus uniquely determining $i$. Based on $i$ he performs the unitary
\be
V_i^{B} = \sum_s (-1)^{b(s,i)} \ket{\varphi_{s, i}} \bra{\varphi_{s, i}},
\label{vunitary}
\ee
which corrects the phase error. Finally, he un-encodes with $U_\cC^{-1}$.

It is important to note that $V_i^{B}$ need not be efficiently implementable,
which is a realistic problem when the code length becomes large (see Discussion). 

\end{proof}


\begin{lemma}
In notation from the proof of Theorem 1,
\be
\frac{1}{2^m}\sum_s \bbE_f\left\|\sigma_s^E(f) - \sigma_0^E(f)\right\|_1 \leq 2\eps',
\ee 
where $\sigma^E_s(f)=\frac{1}{2^{k-m}}\sum_{z\in C'}\phi_{z+x_s}^E$ 
and where $\eps' = 2^{-\frac{1}{2}(H_2(XE)_{\omega} - H_0(E)_{\omega} + k - n - m)}$.
\end{lemma}
\begin{proof}
Consider the state  
\be
\sigma^{YE} = \frac{1}{2^k} \sum_{y\in\bbZ_2^k} \ket{y}\bra{y}^Y \otimes \phi_{Gy}^{E}~.
\label{sigmaYE}
\ee
By Lemma \ref{universal}, we have the privacy condition
\be
\begin{aligned}
\bbE_f \left\|\sigma^{SE}(f) - \tau^{S} \otimes \sigma^E\right\|_1\leq 2^{-\frac{1}{2}(H_2(YE)_{\sigma} - H_0(E)_{\sigma}- m)}~,
\label{PA1}
\end{aligned}
\ee 
with 
$$
\sigma^{SE}(f) = \frac{1}{2^m} \sum_{s\in\bbZ_2^m} \ket{s}\bra{s}^S \otimes \sigma^E_s(f)~,
$$
where 
$$
\sigma^E_s(f)=\frac{1}{2^{k-m}}\sum_{y:f(y)=s}\phi_{Gy}^E=\frac{1}{2^{k-m}}\sum_{z\in C'}\phi_{x_s + z}^E
$$
and $\tau^{S}=\frac{1}{2^m}\sum_s\ket{s}\bra{s}^{S}$ is the maximally mixed state. 

By the  relations (see Appendix E for details)
$H_2(YE)_{\sigma} = H_2(XE)_{\omega} - (n - k)$ and
$H_0(E)_{\sigma} \geq H_0(E)_{\omega}$,
with $\omega$ defined in (\ref{omegaXBE}),
the condition ($\ref{PA1}$) can be easily generalized as 
$$
\bbE_f \left\|\sigma^{SE}(f) - \tau^{S} \otimes \sigma^E\right\|_1 \leq 2^{-\frac{1}{2}(H_2(XE)_{\omega} - H_0(E)_{\omega} + k - n - m)}~. 
$$
This can be rewritten as 
$$
\frac{1}{2^m}\sum_s \bbE_f\left\|\sigma_s^E(f) - \sigma^E\right\|_1 \leq \eps'~,
$$
where $\eps' = 2^{-\frac{1}{2}(H_2(XE)_{\omega} - H_0(E)_{\omega} + k - n - m)}$.
Without loss of generality, we can assume that
$$\bbE_f||\sigma_0^E(f) - \sigma^E||_1\leq\eps',$$ 
and therefore we have
\be
\frac{1}{2^m}\sum_s \bbE_f\left\|\sigma_s^E(f) - \sigma_0^E(f)\right\|_1 \leq 2\eps'~.
\ee
\end{proof}
\vspace{5mm}

\section{Code performance on memoryless qubit Pauli channels}

We now consider the important case of i.i.d. (independent and identically distributed)
or tensor power channels $\cN_n = \hat{\cN}^{\otimes n}$, where $\hat{\cN}$ is 
a single qubit Pauli channel: 
$$
 \hat{\cN}(\rho) = \sum_{u, v \in \bbZ_2}  p_{u,v} X^{u}Z^{v} \rho Z^{v}X^{u}~.
$$
First, we need to characterize the error parameter 
$\eta$ from Theorem 1 using smooth R\'enyi entropy. Recall that 
$$
\eps' = 2^{-\frac{1}{2}(H_2(XE)_{\omega} - H_0(E)_{\omega} + k - n - m)}~,
$$ 
is the error parameter introduced to describe the privacy amplification condition ($\ref{PA2}$). 
Then by Corollary \ref{smooth} (see Appendix), we can replace $\eps'$ by
\be
\eps_1 = 2^{-\frac{1}{2}(H_{\infty}^{\delta}(XE)_{\omega} - H_0^{\delta}(E)_{\omega} + k - n - m)} + 2 \delta~.
\label{eps1}
\ee 
For our i.i.d. case we have $\omega = \hat{\omega}^{\otimes n}$
with
$$
\hat{\omega}^{XBE} = \frac{1}{2}\sum_{x\in \bbZ_2}\ket{x}\bra{x}^X \otimes 
U_{\hat{\cN}}  \ket{{x}}\bra{x} U_{\hat{\cN}}^\dagger~.
$$ 
Then as $n \ra \infty$, by Lemma $\ref{iid}$ we have 
\begin{align*}
\frac{1}{n}H_{\infty}^{\delta}(XE)_{\hat{\omega}^{\otimes n}} &\ra H(XE)_{\hat{\omega}} + o(\delta)~,\\
\frac{1}{n}H_0^{\delta}(E)_{\hat{\omega}^{\otimes n}}  &\ra H(E)_{\hat{\omega}} + o(\delta)~. 
\end{align*} 
Observing $H(X)_{\hat{\omega}} = 1$, then as $n \ra \infty$ we have
$$
\frac{1}{n}H_{\infty}^{\delta}(XE)_{\omega} - \frac{1}{n}H_0^{\delta}(E)_{\omega} -1 \ra - I(X; E)_{\hat{\omega}} + o(\delta)~.
$$

Notice that the state $\omega^{XBE}$ represents the classical-quantum correlation for Alice 
sending classical strings over a Pauli channel. A memoryless Pauli channel 
works as a classical binary symmetric channel for classical information. So there exist good classical 
$[n, k]$ codes such as LDPC codes with rates 
$$
\frac{k}{n} = I(X; B)_{\hat{\omega}} - \Delta 
$$ 
approaching the classical Shannon capacity $$I(X; B)_{\hat{\omega}} = 1 - H(\{p_{00} + p_{01}, p_{10} + p_{11} \})$$
since we have
\begin{align*}
\hat{\omega}^{XB} = \frac{1}{2}\sum_{x\in \bbZ_2}\ket{x}\bra{x}^X \otimes ((p_{00} + p_{01})\proj{x}^B\\
+~ (p_{10} + p_{11})\proj{x+1}^B)~,
\end{align*}
and the error probability $\eps \ra 0$ as $n \ra \infty$, 
where $p$ is the probability of a bit flip error and $\Delta$ is a constant which can be made quite small
~\cite{LDPC,Mackay99,MN96,SS96,RSU2001}. 

Hence, as $n \ra \infty$, the error parameter $\eps_1$ defined by ($\ref{eps1}$) approaches a limit, i.e.   
$$
\eps_1 \ra \eps_2 = 2^{-\frac{1}{2}\{ n[C - \Delta + o(\delta)] - m \}} 
+ 2 \delta~,
$$
where 
$$
C = I(X; B)_{\hat{\omega}} - I(X; E)_{\hat{\omega}} = 1 - H(\{p_{00}, p_{01}, p_{10}, p_{11} \})
$$
is the hashing bound on the Pauli channel \cite{BDSW96}. 
Therefore, the rate of our entanglement generation codes can go up to
$$
\frac{m}{n} = C - \Delta~,
$$
such that $\eps_2 \ra 0$ as $n \ra \infty$ and $\delta \ra 0$. 

Therefore, if we employ LDPC codes as our classical codes, we can get a family of $\eta$-good entanglement 
generation codes with code rate approaching the hashing bound of the memoryless Pauli channels and 
the error parameter $\eta\ra 0$ as $n \ra \infty$. 

Here we want to present an example of P-CSS codes based on an LDPC code from David Mackay's classical 
paper \cite{Mackay99}. The classical code is a Gallager code \cite{LDPC} with $n = 19839$, $k =9839$ 
and each column of its parity check matrix has weight $t = 3$. For practical purpose, we shall extend 
the error parameter $\eps$ defined by ($\ref{parameter3}$) to include both detected and undetected errors 
\footnote{Detected errors occur if the decoder identifies the block to be in the error but its 
algorithm runs for the maximum number of iterations without finding a valid decoding. And the 
original definition $\eps = 1-\sum_ep_{\hat{u}(e)}$ is for undetected errors, which occur if 
the decoder finds a valid decoding that is not the correct decoding.}.
For bit flip error probability equal to $0.076$, the paper gave an estimate of the block error 
probability to be $2.62\times 10^{-5}$, not specified to be detected or undetected errors.
So $2.62\times 10^{-5}$ is an optimistic estimate of total block error probability.  

\begin{figure}
\centerline{ {\scalebox{0.45}{\includegraphics{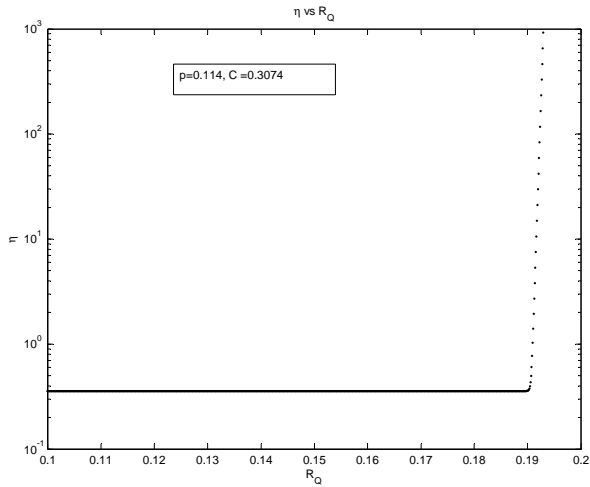}}}}
  \caption{The error parameter $\eta$ vs the quantum code rate $R_Q$ for the Gallager code 
  $n = 19839$, $k =9839$ and column weight $t = 3$.}
\label{Ratecurve}  
\end{figure}

Then we consider the code performance on an iid depolarizing channel 
$$
\hat{\cN}(\rho) = (1-p)\rho+\frac{p}{3}(X\rho X + Y\rho Y + Z\rho Z).
$$
Since $n$ is large enough, we can estimate $\eps'$ by
$\eps'= 2^{-\frac{1}{2}(-nI(X;E)_{\hat{\omega}} + k - m)}$ and for $p=0.076*3/2=0.114$,
we have $I(X;E)_{\hat{\omega}} = 0.3046$, $C = 0.3074$.
By setting $R_Q=m/n$, we can plot $\eta$ vs $R_Q$ in Fig.\ref{Ratecurve}. 
We see that $\eta$ stabilizes at $0.3548$ right after $R_Q<0.19$. 
Note that $\eta$ is a rough upper-bound for the real error, which is believed to 
be much smaller than $0.3548$. So Fig.\ref{Ratecurve} is just for illustration 
purpose and it should not be considered as the real code performance.
 
\section{Discussion}

We have studied a subclass of CSS codes called P-CSS codes, which are based on a classical error correcting code and a 
two-universal hash function. P-CSS codes are very flexible and easy to construct, 
and have excellent asymptotic performance.
However, there are several drawbacks, relative to the traditional finite-distance CSS codes, that deserve further study.
The phase-flip correcting unitary operation $V_i$ from (\ref{vunitary}) suffers from two kinds of
inefficiencies. First, it is not a tensor power of single qubit operations, but a generic $n$-qubit 
operation. Contrast this to the usual way of correcting phase-flip errors: given the error syndrome 
$i \in \bbZ_2^{k - m}$, one associates to it the most probable error $\hat{v}(i) \in \bbZ_2^{n}$ (cf. the function $\hat{u}(e)$ for bit-flip errors), and performs $Z^{\hat{v}(i)}$. In particular there is no $s$-dependence
as in $b(s,i)$ from (\ref{vunitary}). Second, even if the phase-flip correcting unitary is of the form $Z^{\hat{v}(i)}$,
the function $\hat{v}(i)$ may not be efficiently computable. This issue is not present for $\hat{u}(e)$ if we
use an efficiently decodable LDPC code for $C$. We believe that at least the first issue can be overcome by 
modifying the  Theorem 1 proof technique. Finally, the error parameter $\eta$ is a rough upper-bound of 
the real error and a better bound will be highly desired for applications.

{\bf Acknowledgments} \, We thank I. Devetak for guidance and help with the manuscript.
Thanks also go to T. Brun and M. Wilde for comments on the manuscript.
This work was supported by the NSF grant CCF-0524811.
\appendix 

\section{Fidelity and trace distance}
 
It is necessary to recall some facts about trace distances, fidelities, and purifications 
(mostly taken from \cite{NC00}). The trace distance between two density operators $\rho$ and $\sigma$
can be defined as  
$$||\rho - \sigma||_1 = \tr|\rho - \sigma|,$$
where $|A|\equiv\sqrt{A^{\dagger}A}$ is the positive square root of $A^{\dagger}A$.
The fidelity of two density operators with respect to each other can be defined as
$$F(\rho, \sigma) = \| \sqrt{\rho} \sqrt{\sigma} \|^2_1.$$
For two pure states $\ket{\chi}$, $\ket{\zeta}$ this
amounts to
$$
F(\ket{\chi}, \, \ket{\zeta}) = |\langle \chi \ket{\zeta}|^2.
$$

The following relation between fidelity and trace distance will be needed:
\be
1 - \sqrt{F(\rho, \sigma)} \leq  \frac{1}{2} \| \rho - \sigma \|_1
\leq \sqrt{1 - F(\rho, \sigma)},\label{tracefidelity}
\ee
the second inequality becoming an equality for pure states.
\vspace{1mm}

A \emph{purification}  $\ket{\Phi_\rho}^{RB}$ of a density operator $\rho^B$ is some pure
 state living in an augmented quantum system $RB$ such that 
 $\tr_{R}( \ket{\Phi_\rho}\bra{\Phi_\rho}^{RB}) = \rho^B$. Any two purifications 
 $\ket{\Phi_\rho}^{RB}$ and $\ket{\Phi'_\rho}^{RB}$ of $\rho^B$ are related by 
some local unitary $U$ on the \emph{reference system} $R$
$$
\ket{\Phi'_\rho}^{RB} = (U^{R}  \otimes I^{B}) \ket{\Phi_\rho}^{RB}.
$$
A theorem by Uhlmann states that, for a fixed purification $\Phi_\sigma$ of $\sigma$,
$$
F(\rho, \sigma)  = \max_{\Phi_\rho} F(\ket{\Phi_\rho}, \, \ket{\Phi_\sigma}).
$$
A corollary of this theorem is the \emph{monotonicity} property of fidelity 
\be
F(\rho^{RB}, \sigma^{RB}) \leq F(\rho^{B}, \sigma^{B}),
\label{monotonicity}
\ee 
where $\rho^B =  \tr_{R} \rho^{RB}$ and $\sigma^B =  \tr_{R} \sigma^{RB}$. 
The corresponding \emph{monotonicity} property of trace distance is 
\be
||\rho^{RB} - \sigma^{RB}||_1 \geq ||\rho^{B} - \sigma^{B}||_1~.
\label{monodis}
\ee
Another important property of fidelity is \emph{concavity}
\be
F\left( \sum_i p_i \rho_i, \sum_i q_i \sigma_i \right) \geq \left(\sum_i \sqrt{p_iq_i} \sqrt{F(\rho_i, \sigma_i)}\right)^2,
\label{concavity}
\ee
where $p_i \geq 0$, $q_i \geq 0$, and $\sum_i p_i = \sum_i q_i = 1$.



\section{R\'enyi entropy} 

The following definitions and properties of R\'enyi entropy are mostly taken from~\cite{RK04}.  
For $\alpha\in[0, \infty]$ and a density operator $\rho$, the \emph{R\'enyi entropy of order $\alpha$ 
of $\rho$} is defined by
$$
H_{\alpha}(\rho) := \frac{1}{1-\alpha}\log_2(\tr(\rho^{\alpha}))~, 
$$
with the convention $H_{\alpha}(\rho) := \lim_{\beta\ra\alpha}H_{\beta}(\rho)$ for $\alpha\in\{0,1,\infty\}$.

In particular, for $\alpha = 0$, $H_0(\rho) = \log_2(\rank(\rho))$; for $\alpha = 1$, $H_1(\rho)$ is the 
von Neumann entropy $H(\rho)$; for $\alpha = \infty$, $H_{\infty}(\rho) = -\log_2(\lambda_{\max}(\rho))$, 
where $\lambda_{\max}(\rho)$ denotes the maximum eigenvalue of $\rho$. Furthermore, for $\alpha$, 
$\beta \in [0, \infty]$,
$$
\alpha \leq \beta  \,\Leftrightarrow\, H_{\alpha}(\rho) \geq H_{\beta}(\rho)~. 
$$

The definition of R\'enyi entropy for density operators can be generalized to the notion of smooth R\'enyi 
entropy. For $\alpha\in[0, \infty]$, $\eps\geq 0$ and a density operator $\rho$, the \emph{$\eps$-smooth R\'enyi entropy of order $\alpha$ of $\rho$} is defined by
\begin{align*}
H_{\alpha}^{\eps}(\rho) := 
\left\{
     \begin{aligned}  
     &\inf_{\sigma\in\cB^{\eps}(\rho)} H_{\alpha}(\sigma), \,\,\,0 \leq \alpha < 1  \\
     &\sup_{\sigma\in\cB^{\eps}(\rho)} H_{\alpha}(\sigma), \,\,\,1 < \alpha \leq \infty
     \end{aligned}
\right.     
\end{align*}
where $\cB^{\eps}(\rho) = \{\sigma : ||\rho - \sigma||_1 \leq \eps\}$ and $H_{1}^{\eps}(\rho) = H(\rho)$.

In the independent and identically distributed (i.i.d.) case, the smooth R\'enyi entropy $\frac{1}{n}H_{\alpha}^{\eps}(\rho^{\otimes n})$ of the state $\rho^{\otimes n}$ will equal 
its Shannon entropy $H(\rho)$ as $n$ goes to infinity.

\begin{lemma}\label{iid} 
For a density operator and any $\alpha \in [0, \infty]$, 
\be
\lim_{\eps\ra 0}\lim_{n\ra\infty}\frac{H_{\alpha}^{\eps}(\rho^{\otimes n})}{n} = H(\rho),\,\, 
\forall \alpha\in[0, \infty]~.
\ee
\end{lemma}

\section{Universally composable privacy amplification}

Now we are ready to introduce an important lemma by Renner and K\"onig~\cite{RK04}.
\begin{lemma}\label{universal}
For a classical-quantum system $YE$ with state $\sigma^{YE} = \sum_{y\in\cY} p(y)\ket{y}\bra{y}^{Y}
\otimes \sigma^E_y$, let $f$ be a two-universal function on $\cY$ with range $\bbZ_2^{m}$, which is 
independent of $YE$. Then
$$
\bbE_f \left\|\sum_{s}q(s|f)\ket{s}\bra{s}^{S} \otimes \sigma_s^E(f) - \tau^{S} \otimes \sigma^E\right\|_1 
\leq \eps~, 
$$
where 
$$
\eps = 2^{-\frac{1}{2}(H_2(YE)_{\sigma} - H_0(E)_{\sigma}-m)}~,
$$ 
$S = f(Y)$ with probability distribution $q$, $\sigma^E_s(f) = \frac{1}{q(s|f)}\sum_{y\in f^{-1}(s)}p(y)\sigma_y^E$ with $f^{-1}(s) = \{y | f(y) = s\}$, and $\tau^{S} = \frac{1}{2^m}\sum_s\ket{s}\bra{s}^{S}$ is the maximally mixed state.
\end{lemma}

The result of Lemma $\ref{universal}$ can be generalized to smooth R\'enyi entropy.

\begin{corollary}\label{smooth}
For a classical-quantum system $YE$ with state $\sigma^{YE} = \sum_{y\in\cY} p(y)\ket{y}\bra{y}^{Y}
\otimes \sigma^E_y$, let $f$ be a two-universal function on $\cY$ with range $\bbZ_2^{m}$, which is 
independent of $YE$. Then for $\eps\geq 0$
$$
\bbE_f \left\|\sum_{s}q(s|f)\ket{s}\bra{s}^{S} \otimes \sigma_s^E(f) - \tau^{S} \otimes \sigma^E\right\|_1 
\leq \eps'~, 
$$
where 
$$
\eps' = 2^{-\frac{1}{2}(H_{\infty}^{\eps}(YE)_{\sigma} - H_0^{\eps}(E)_{\sigma}-m)} + 2\eps~,
$$ 
$S = f(Y)$ with probability distribution $q$, $\sigma^E_s(f) = \frac{1}{q(s|f)}\sum_{y\in f^{-1}(s)}p(y)\sigma_y^E$ with $f^{-1}(s) = \{y | f(y) = s\}$, and $\tau^{S} = \frac{1}{2^m}\sum_s\ket{s}\bra{s}^{S}$ is the maximally mixed state.
\end{corollary}

\medskip

\section{Evaluating R\'enyi entropy for Pauli channels}
Given the expression ($\ref{standard}$) of $\ket{\phi_x}^{BE}$, we have
$$
\phi_x^E = \sum_{u} p_u \ket{u}\bra{u}^{E_1} \otimes \ket{\phi_{x, u}}\bra{\phi_{x, u}}^{E_2}~,
$$ 
with $\ket{\phi_{x, u}}^{E_2} = \sum_v \sqrt{p_{v|u}} (-1)^{v\cdot x} \ket{v}^{E_2}$. 

Observe that $\tr(\phi_x^E)^2 = \sum_{u } p_u^2$ is independent of $X$. Then for 
$\omega^{XBE}$ defined by ($\ref{omegaXBE}$) and $\sigma^{YE}$ defined by ($\ref{sigmaYE}$),
we have
\begin{align*}
H_2(YE)_{\sigma} &= k - \log_2\left(\sum_{u} p_u^2\right)~,\\
H_2(XE)_{\omega}   &= n - \log_2\left(\sum_{u} p_u^2\right)~.
\end{align*} 
So $H_2(XE)_{\omega} = n - k + H_2(YE)_{\sigma}$. 
\medskip

To show $H_0(E)_{\omega}\geq H_0(E)_{\sigma}$,
we need to introduce \emph{Weyl's monotonicity theorem}~\cite{Bhatia97}. 
For a Hermitian operator $A$, define
$\lambda^{\downarrow}(A) = (\lambda^{\downarrow}_1(A),\ldots,\lambda^{\downarrow}_n(A))$, where eigenvalues $\lambda^{\downarrow}_j(A)$ are arranged in decreasing order.

\begin{thm}[Weyl's monotonicity theorem]
If $A$ is Hermitian and $B$ is positive, then for all $j$
$$
\lambda^{\downarrow}_j(A + B) \geq \lambda^{\downarrow}_j(A)~.
$$
\end{thm}

Then if both $A$ and $B$ are density operators, the number of positive eigenvalues of $A + B$ should be
no less than that of $A$, i.e. $\text{rank}(A + B) \geq \text{rank}(A)$. 
Note the definition of $\sigma^{XE}$ can be generalized to $\sigma_\ell^{XE}$ by changing the classical 
code $C$ to the $\ell$th coset of $C$ in $\bbZ_2^n$. Given $\omega^{E} = \frac{1}{2^{n - k}}\sum_{\ell} \sigma_\ell^{E}$, we have $\text{rank}(\omega^E) \geq \text{rank}(\sigma^E_\ell)$, i.e. $H_0(E)_{\omega} \geq H_0(E)_{\sigma_\ell}$ for all $\ell$.
  

\bibliography{ref}
\bibliographystyle{abbrv}

\end{document}